\documentclass[12pt]{article}
\usepackage{epsfig,amsfonts,amssymb}
\usepackage{cite}
\input epsf.sty
\usepackage{cite}

\def\ZZZ{{\hbox{ Z\kern-1.6mm Z}}}
\def\RRR{{\hbox{ R\kern-2.4mm R}}}
\def\CCC{{\hbox{ C\kern-2.0mm C}}}
\def\zzz{{\hbox{z\kern-1mm z}}}

\newcommand{\qeq}{{\hbox{=\kern-2.3mm ? \kern.5mm }}}
\renewcommand{\qeq}{=}

\newcommand{\NN}{{\cal N}}

\newcommand{\be}{\begin{equation}}
\newcommand{\ee}{\end{equation}}
\newcommand{\ben}{\begin{eqnarray}\displaystyle}
\newcommand{\een}{\end{eqnarray}}

\newcommand{\bea}[1]{\begin{eqnarray}\label{#1} }
\newcommand{\eea}{\end{eqnarray}}

\newcommand{\refb}[1]{(\ref{#1})}

\def\one{{\hbox{ 1\kern-.8mm l}}}
\def\zero{{\hbox{ 0\kern-1.5mm 0}}}

\begin{document}

{}~
{}~

\vskip .6cm

{\baselineskip20pt
\begin{center}
{\Large \bf
Rare Decay Modes of Quarter BPS Dyons} 

\end{center} }

\vskip .6cm
\medskip

\vspace*{4.0ex}

\centerline{\large \rm
Ashoke Sen}

\vspace*{4.0ex}

\centerline{\large \it Harish-Chandra Research Institute}

\centerline{\large \it  Chhatnag Road, Jhusi,
Allahabad 211019, INDIA}

\vspace*{1.0ex}

\centerline{E-mail: sen@mri.ernet.in, ashokesen1999@gmail.com}

\vspace*{5.0ex}

\centerline{\bf Abstract} \bigskip

 The degeneracy of quarter BPS dyons in N=4 supersymmetric
 string theories is known to jump across
walls of marginal stability on which a quarter
BPS dyon can decay into a pair of half BPS dyons. We show that as
long as the electric and magnetic charges of the original dyon are
primitive elements of the charge lattice, the
subspaces of the moduli space on which a quarter BPS dyon becomes
marginally
unstable against decay into a pair of quarter BPS dyons or a
half BPS dyon and a quarter BPS dyon are of codimension two or
more. As a result any pair of generic points  in the
moduli space can be connected by a path
avoiding these subspaces and there is no jump in
the spectrum associated with these subspaces.

\vfill \eject

\baselineskip=18pt

We now have a good understanding
of the exact spectrum of a class of
quarter BPS dyons in $\NN=4$
supersymmetric string 
theories\cite{9607026,0412287,
0505094,0506249,0510147,0602254,
0603066,0605210,0607155,0609109,0612011,0702141,
0702150,0705.1433,0705.3874,0706.2363}. 
It is also known that as we cross
various walls of marginal stability associated with the possible
decay of the dyon into a pair of half BPS states,
the degeneracy changes by a
certain amount that is exactly 
computable\cite{0702141,0702150}.
Furthermore in the gravity description this jump 
can be accounted for by the (dis)appearance of two
centered small black 
holes\cite{0005049,0010222,0101135,0304094,0702146,0705.2564} 
as in the asymptotic moduli space
we cross walls of marginal 
stability\cite{0702150,0705.3874,0706.2363}.
This raises the question:
why aren't there similar effects associated with the decay of
a quarter BPS dyon into a pair of quarter BPS dyons, or into a
quarter BPS dyon and a half BPS dyon?
In this note we
shall show that
such decays take place on subspaces of codimension higher
than one as long as electric and magnetic charge vectors 
of the original
dyon are primitive elements of the charge lattice. 
Hence we can move from any generic point 
in the moduli space to another generic point in the moduli space
without ever passing through these subspaces, and there is no effect
of the type discussed in \cite{0702141,0702150,0705.3874,0706.2363}
associated with these 
decays.\footnote{Different 
approaches to this problem 
have been advocated in
\cite{denef,cheng}.}

We denote by $r$ the total number of U(1) gauge fields in the
model, by $\vec Q$ and 
$\vec P$ the $r$ dimensional
electric and the magnetic charge vectors,
by $\tau=a+iS$ the axion-dilaton moduli field parametrizing the
upper half plane and
by $M$ the $r\times r$  matrix valued scalar field satisfying
\be \label{emcon}
MLM^T=L, \qquad M^T=M, \qquad L=
\pmatrix{I_6 &\cr & -I_{r-6}}\, 
\ee
where $I_k$ denotes $k\times k$ identity matrix. 
We shall use the subscript $_\infty$ to denote
the asymptotic values of various scalar fields. Let us now
introduce the $SO(6,r-6)$ matrix $\Omega_\infty$
via the relations\footnote{Since \refb{edefomin} is invariant
under a right multiplication of $\Omega_\infty$ by an $SO(6)\times
SO(r-6)$
matrix that preserves both the identity matrix and $L$, 
\refb{edefomin} does not fix $\Omega_\infty$ completely in terms
of $M_\infty$. This problem may be avoided by choosing a suitable
`gauge condition' on $\Omega_\infty$ so that there is one to
one correspondence between $M_\infty$ and $\Omega_\infty$.}
\be \label{edefomin}
M_\infty =\Omega_\infty \Omega_\infty^T, 
\qquad
\Omega_\infty L \Omega_\infty^T=L \, ,
\ee
and define
\be \label{edefqrpr}
Q_R ={1\over 2} (I_r + L) \Omega_\infty^T\, Q, \qquad
P_R ={1\over 2} (I_r + L) \Omega_\infty^T\,  P\, .
\ee
The vectors $\vec Q_R$ and $\vec P_R$
lie in the six dimentional subspace spanned by the eigenvectors of $L$
with eigenvalue 1. 
In terms of $\vec Q_R$ and $\vec P_R$
the BPS mass formula of 
\cite{9507090,9508094} 
takes the form\cite{0702141}, 
\be \label{emay}
m(\vec Q, \vec P) = \sqrt{2}\,
f(\vec Q_R, \vec P_R; a_\infty, S_\infty)\, ,
\ee
where 
\be \label{emay2}
f(\vec Q_R, \vec P_R; a_\infty, S_\infty)
=\sqrt{{1\over S_\infty} 
(\vec Q_R - a_\infty \vec P_R)^2 + S_\infty \vec P_R^2
+ 2 \left[ \vec Q_R^2 \vec P_R^2 -
( \vec Q_R\cdot  \vec P_R)^2\right]^{1/2}}\, .
\ee
The inner products of $\vec Q_R$ and $\vec P_R$ are calculated
with the identity matrix.

Let us now consider a possible marginal
decay $(\vec Q,\vec P)\to
(\vec Q_1, \vec P_1) + (\vec Q-\vec Q_1, \vec P - \vec P_1)$.
This requires adjusting the moduli such that
\be \label{eadjust}
f(\vec Q_R, \vec P_R; a_\infty, S_\infty)
=f(\vec Q_{1R}, \vec P_{1R}; a_\infty, S_\infty)
+ f(\vec Q_R-\vec Q_{1R}, \vec P_R-
\vec P_{1R}; a_\infty, S_\infty)\, .
\ee
For fixed $\vec Q$, $\vec P$ and $M_\infty$,
$\vec Q_R$ and $\vec P_R$ span a two dimensional subspace
of the six dimensional space on which $L$ has eigenvalue +1.
Let us denote by $\vec Q_{1R\parallel}$ and 
$\vec P_{1 R\parallel}$ the projection of $\vec Q_{1R}$ and 
$\vec P_{1R}$ along this two dimensional subspace. Then we
have the following inequalities:
\be \label{ineq1}
f(\vec Q_R, \vec P_R; a_\infty, S_\infty)
\le f(\vec Q_{1R\parallel}, \vec P_{1R\parallel}; a_\infty, S_\infty)
+ f(\vec Q_R-\vec Q_{1R\parallel}, \vec P_R-
\vec P_{1R\parallel}; a_\infty, S_\infty)\, ,
\ee
\be \label{ineq2}
f(\vec Q_{1R\parallel}, \vec P_{1R\parallel}; a_\infty, S_\infty)
\le f(\vec Q_{1R}, \vec P_{1R}; a_\infty, S_\infty)\, ,
\ee
\be \label{ineq3}
f(\vec Q_R-\vec Q_{1R\parallel}, \vec P_R-
\vec P_{1R\parallel}; a_\infty, S_\infty)
\le 
f(\vec Q_R-\vec Q_{1R}, \vec P_R-
\vec P_{1R}; a_\infty, S_\infty)\, .
\ee
The inequality \refb{ineq1} is proved by defining
\be \label{epr1}
\vec a = {\vec Q_R - a_\infty \vec P_R\over \sqrt{S_\infty}}, \qquad
\vec b = {\vec P_R\, \sqrt{S_\infty}}\, ,
\qquad
\vec a_1 = {\vec Q_{1R\parallel} - a_\infty \vec P_{1R\parallel}
\over \sqrt{S_\infty}}, \qquad
\vec b_1 = {\vec P_{1R\parallel}\, \sqrt{S_\infty}}\, ,
\ee
and using the inequality:
\be \label{epr2}
\sqrt{\vec a^2 + \vec b^2 + 2|\vec a\times \vec b|}
\le \sqrt{\vec a_1^2 + \vec b_1^2 + 2|\vec a_1\times \vec b_1|}
+ \sqrt{(\vec a-\vec a_1)^2 + (\vec b-\vec b_1)^2 + 2|(\vec a
-\vec a_1)\times (\vec b-\vec b_1)|}\, 
\ee
for any set of
vectors $\vec a$, $\vec b$, $\vec a_1$, $\vec b_1$ lying in a
two dimensional plane. \refb{epr2} can be easily proven with the
help of triangle inequality if we note that 
$\sqrt{\vec a^2 + \vec b^2 + 2|\vec a\times \vec b|}$ can be
interpreted as $|\vec a + \epsilon
\vec b|$ where $\epsilon$ is the $\pi/2$ rotation matrix in the
plane of $\vec a$ and $\vec b$, with the sign of $\epsilon$ chosen
such that $a^T\epsilon b>0$.
Requiring the inequality \refb{epr2} to be saturated
gives one equation and several inequalities
among the components of $\vec a$, $\vec b$,
$\vec a_1$ and $\vec b_1$:
\be \label{einco}
\vec a_1 +\epsilon \vec 
b_1=\lambda (\vec a +
\epsilon\vec b)\quad \hbox{with} 
\quad 0\le \lambda\le 1\, , \qquad
a_1^T \epsilon b_1\ge0, \qquad (a-a_1)^T \epsilon (b-b_1)\ge0
\, .
\ee
Using \refb{epr1} we can
translate these conditions into one constraint equation
and some inequalities 
involving the variables
$(a_\infty, S_\infty,M_\infty,\vec Q, \vec P)$.

The inequality \refb{ineq2} follows from the observations that
\ben \label{eobs1}
&& |\vec Q_{1R\parallel} - \bar\tau_\infty \vec P_{1R\parallel}|^2 \le
|\vec Q_{1R} - \bar\tau_\infty \vec P_{1R}|^2, \nonumber \\
&&
\sqrt{ \vec Q_{1R\parallel}^2 \vec P_{1R\parallel}^2 -
( \vec Q_{1R\parallel}\cdot  
\vec P_{1R\parallel})^2} \le 
\sqrt{ \vec Q_{1R}^2 \vec P_{1R}^2 -
( \vec Q_{1R}\cdot  \vec P_{1R})^2}\, .
\een
The first of these is obvious since the (complex)
vector on the left hand side
is a projection of the vector on the right hand side along
the plane spanned by $\vec Q_R$ and $\vec P_R$. 
The second one follows from the fact that
the right hand side of the inequality represents the area of a triangle
formed by the vectors $\vec Q_{1R}$ and $\vec P_{1R}$ and the
left hand side represents the area of the projection of this triangle in
the plane spanned by $\vec Q_R$ and $\vec P_R$. Both inequalities
are saturated when $\vec Q_{1R}$ and $\vec P_{1R}$ lie in the
plane spanned by $\vec Q_R$ and $\vec P_R$. The inequality
\refb{ineq3} can be proved in an identical manner and is also
saturated when $\vec Q_{1R}$ and $\vec P_{1R}$ lie in the
plane spanned by $\vec Q_R$ and $\vec P_R$. This requires
adjusting $\Omega_\infty$ or equivalently $M_\infty$ appropriately.

Now in order to satisfy the condition for marginal stability
\refb{eadjust} we must saturate all the three inequalities
\refb{ineq1}-\refb{ineq3}. This would require adjusting moduli
$M_\infty$ to make $(\vec Q_{1R}, \vec P_{1R})$ lie in the plane
of $(\vec Q_R, \vec P_R)$, and additional adjustment of
$(a_\infty,S_\infty)$ to saturate the inequality \refb{ineq1}. Thus
we have a surface of codimension two or more, and we
can go from any generic point in the moduli space to another
generic point in the moduli space without ever encountering this
subspace of marginal stability. This shows that there is no
discontinuous change in the spectrum associated with these
subspaces.

It is instructive to compare this with the condition for marginal
stability of half BPS dyons in the $\NN=2$ supersymmetric
S-T-U model. In that case the BPS mass formula is identical to the
one given in \refb{emay}, \refb{emay2}, but $M$ and $L$ are
$4\times 4$ matrices and $L$ has two eigenvalues +1 and 
two eigenvalues $-1$. As a result the vectors $\vec Q_R$, $\vec P_R$,
$\vec Q_{1R}$, $\vec P_{1R}$ all lie inside a two dimensional
subspace spanned by the eigenvectors of $L$ with eigenvalue
+1, and the inequalities \refb{ineq2}, \refb{ineq3} are
automatically saturated. Thus we only need to saturate the
inequality \refb{ineq1}. This gives one condition on the
asymptotic moduli, producing a codimension one surface.

There is a special case where our argument fails for the $\NN=4$
supersymmetric theory. If the full $r$ dimensional
charge vectors $\vec Q_1$ and
$\vec P_1$ happen to lie in the plane spanned by $\vec Q$ and $\vec P$
then $\vec Q_{1R}$ and $\vec P_{1R}$ automatically lie in the
plane of $\vec Q_R$ and $\vec P_R$ and we do not get any
condition on the moduli $M_\infty$ from
\refb{ineq2}, \refb{ineq3}. This would require $\vec Q_1$
and $\vec P_1$ to be of the form:
\be \label{eform1}
\vec Q_1=\alpha \vec Q + \beta \vec P, \qquad \vec P_1=
\gamma \vec Q+\delta \vec P\, .
\ee
If we take $\vec Q$ and $\vec P$ to be primitive then charge
quantization would require $\alpha$, $\beta$, $\gamma$, $\delta$
to be integers. Furthermore in the $\ZZZ_N$ orbifold models
$\gamma$ must be integer multiples of $N$ since in one
particular direction along the charge lattice
$Q$ is quantized in units of $1/N$ while $P$ is quantized in integer
units\cite{0510147,0605210}. 
\refb{eform1} now implies that
\be \label{eform2}
\vec Q_{1R}=\alpha \vec Q_R + \beta \vec P_R, \qquad \vec 
P_{1R}=
\gamma \vec Q_R+\delta \vec P_R\, .
\ee
We can substitute these into \refb{ineq1} and use the discussion
below \refb{epr2} to determine under what condition the
inequality might be saturated. 
We shall only 
consider the case when $\vec Q_R$ and $\vec P_R$ are
not parallel, \i.e.\ $\vec Q_R^2 \vec P_R^2-(\vec Q_R\cdot\vec P_R)^2
\ne 0$, since for generic $\vec Q$ and $\vec P$ aligning $\vec Q_R$
and $\vec P_R$ will impose more than one condition on $\Omega_\infty$
and will produce a surface of codimension higher than one. In this case
the $a_1^T\epsilon b_1\ge 0$,
$(a-a_1)^T\epsilon (b-b_1)\ge 0$
conditions give
\be \label{efi1}
\alpha \delta - \beta \gamma \ge 0\, , 
\qquad (1-\alpha) (1-\delta) -\beta \gamma \ge 0\, .
\ee
On the other hand the $(\vec a_1+\epsilon\vec b_1)=\lambda
(\vec a +\epsilon \vec b)$ condition gives
\be \label{egi1}
(\alpha-\lambda - a_\infty \gamma) 
\sqrt{\vec Q_R^2 \vec P_R^2 - (\vec Q_R\cdot \vec P_R)^2} +
+ S_\infty \left( \gamma \vec
Q_R\cdot \vec P_R + (\delta-\lambda)
\vec P_R^2\right) =0\, ,
\ee 
\be \label{egi2}
(\beta - (\delta -\lambda) a_\infty ) 
\sqrt{\vec Q_R^2 \vec P_R^2 - (\vec Q_R\cdot \vec P_R)^2} -
S_\infty \left(\gamma \vec Q_R^2 + (\delta-\lambda)
\vec Q_R\cdot \vec P_R\right) = 0\, ,
\ee
where $\lambda$ is an arbitrary parameter with 
\be \label{egi4}
0\le \lambda\le 1\, .
\ee
We can solve for $a_\infty$ using \refb{egi1}
and substitute into \refb{egi2} to get
\be \label{efi3}
\left(\beta\gamma - \alpha\delta + \lambda (\alpha+\delta)
-\lambda^2\right) \sqrt{\vec Q_R^2 \vec P_R^2 - 
(\vec Q_R\cdot \vec P_R)^2}
- S_\infty  \left\{ (\lambda -\delta) \vec P_R 
-\gamma \vec Q_R
\right\}^2 = 0\, .
\ee
If $\gamma=0$ then \refb{egi2} contains additional information
beyond what can be obtained from \refb{efi3} and \refb{egi1}.

We focus on the constraint \refb{efi3}. Since the second term proportional
to $S_\infty$ is negative semi-definite, 
if we can show that the first term is
also negative definite then we would have shown that the equation has
no solution. For this analysis we shall make use of \refb{efi1}. Let
us consider the following cases separately:
\begin{itemize}
\item First consider the case when both the inequalities in \refb{efi1}
are saturated: 
\be \label{efi4}
\alpha \delta - \beta \gamma = 0\, , 
\qquad (1-\alpha) (1-\delta) -\beta \gamma = 0\, .
\ee
This correponds to the case when in each of the decay products
the electric and the magnetic charge vectors are parallel.
Hence both the decay products are
half BPS. \refb{efi4} gives $\alpha+\delta = 1$. Using this we can
express the first term on the left hand side of \refb{efi3} as
\be \label{efi5}
\left(\lambda
-\lambda^2\right) \sqrt{\vec Q_R^2 \vec P_R^2 - 
(\vec Q_R\cdot \vec P_R)^2}\, .
\ee
Since this is positive semi-definite in the range
\refb{egi4} this can cancel the second term in
\refb{efi3} on a codimension 1 subspace in the $(\lambda, S_\infty)$
space. Using \refb{egi1} or \refb{egi2} 
we can convert this to a codimension 1
subspace in the $(a_\infty,S_\infty)$ space, reproducing the
marginal stability walls studied in \cite{0702141}.
\item Now consider the case where at least one of the decay products
is quarter BPS. Without any loss of generality we can take this to be
the state carrying charges $(\alpha \vec Q+\beta \vec P,
\gamma\vec Q+\delta\vec P)$. In this case the first inequality in
\refb{efi1} will be a strict inequality. Combining this with the
information that $\alpha,\beta,\gamma,\delta$ are integers we get
\be \label{efi6}
\alpha \delta -\beta\gamma\ge 1\, , \qquad 
(1-\alpha) (1-\delta) -\beta \gamma \ge 0\, .
\ee
Our goal is to use these results to analyze the first term on the left
hand side of \refb{efi3}.
Due to \refb{efi6},
\be \label{efi8}
\left(\beta\gamma - \alpha\delta + \lambda (\alpha+\delta)
-\lambda^2\right) 
\ee
is negative or zero at $\lambda=0,1$.
Thus in order for it to be positive in some range of value between
$\lambda=0$ and $\lambda=1$, it must have a maximum in this range,
and its value at the maximum must be positive. Now \refb{efi8} 
has a maximum at
\be \label{efi9}
\lambda = {\alpha +\delta\over 2}\, ,
\ee
where it takes the value
\be \label{efi10}
{1\over 4} (\alpha + \delta)^2 - (\alpha\delta - \beta\gamma)\, .
\ee
Eq.\refb{efi9} shows that 
in order that the maximum lies in the range $(0,1)$ we must have
\be \label{efi11}
0\le (\alpha+\delta) \le 2\, .
\ee
Using eqs.\refb{efi6},
\refb{efi11} we see that \refb{efi10}, representing 
the maximum value of 
\refb{efi8}, must be negative or zero. As a result there is no
cancellation between the two terms in the left hand side of \refb{efi3}.
The only possibility is that both terms may vanish
simultaneously. Vanishing of the second term will require
\be \label{esecond}
\gamma=0, \qquad \lambda = \delta\, ,
\ee
while from \refb{efi6},
\refb{efi9}-\refb{efi11} we see that
the vanishing of the first term would require
\be \label{ethird}
\alpha\delta-\beta\gamma=1,
\qquad \alpha+\delta=2, \qquad \lambda={\alpha+\delta\over 2}\, .
\ee
As a consequence of
\refb{esecond}, \refb{ethird} we get
\be \label{efourth}
\alpha=\delta=1, \qquad \gamma=0, \qquad \lambda=1\, .
\ee
We have seen however that for $\gamma=0$,
\refb{egi1}, \refb{egi2} may contain additional information
beyond the ones which have been already discussed.
In particular substituting \refb{efourth} into \refb{egi2}
we get $\beta=0$. This choice of $(\alpha,\beta,\gamma,\delta)$ 
corresponds to the trivial case where the final decay products 
have charges
$(\vec Q,\vec P)$ and $(0,0)$.

Similar analysis shows that the decay of a quarter BPS dyon into
three or more quarter or half BPS dyons occur on subspaces of
codimension larger than one, since this would require aligning
multiple six dimensional vectors along a plane and/or aligning
multiple two dimensional vectors along a line. 
This completes our proof that the only possible codimension one 
subspaces of marginal stability arise from the decay of a quarter BPS
dyon into a pair of half BPS dyons.

Before concluding this paper we would like to offer a physical
explanation of why decay of a quarter BPS  state into quarter BPS
states requires more constraint than decay into half BPS states.
This essentially arises from the fact that at a point of marginal stability
the supersymmetries of decay products must align. Since half-BPS
states have more supersymmetry than quarter BPS states, it is clearly
easier to ensure that a pair of half BPS states have one common
supersymmetry than ensuring that a pair of quarter BPS states have
a common supersymmetry. Similar argument can be given for the decay
of a quarter BPS state into three or more states.

\end{itemize}

\bigskip

{\bf Acknowledgement:} I would like to thank A.~Mukherjee, 
S.~Mukhi and R.~Nigam for pointing out a logical error in an earlier
version of the draft.


\begin{thebibliography}{99}

  
\bibitem{9607026}
R.~Dijkgraaf, E.~P.~Verlinde and H.~L.~Verlinde,
``Counting dyons in N = 4 string theory,''
Nucl.\ Phys.\ B {\bf 484}, 543 (1997)
[arXiv:hep-th/9607026].

\bibitem{0412287}
  G.~Lopes Cardoso, B.~de Wit, J.~Kappeli and T.~Mohaupt,
  ``Asymptotic degeneracy of dyonic 
  N = 4 string states and black hole
  entropy,''
  JHEP {\bf 0412}, 075 (2004)
  [arXiv:hep-th/0412287].

\bibitem{0505094}
D.~Shih, A.~Strominger and X.~Yin,
``Recounting dyons in N = 4 string theory,''
arXiv:hep-th/0505094.

\bibitem{0506249}
D.~Gaiotto,
``Re-recounting dyons in N = 4 string theory,''
arXiv:hep-th/0506249.

\bibitem{0510147}
  D.~P.~Jatkar and A.~Sen,
  ``Dyon spectrum in CHL models,''
  JHEP {\bf 0604}, 018 (2006)
  [arXiv:hep-th/0510147].

\bibitem{0602254}
  J.~R.~David, D.~P.~Jatkar and A.~Sen,
  ``Product representation of dyon partition function in CHL models,''
  JHEP {\bf 0606}, 064 (2006)
  [arXiv:hep-th/0602254].
  
\bibitem{0603066}
  A.~Dabholkar and S.~Nampuri,  
  ``Spectrum of dyons and black holes in 
  CHL orbifolds using Borcherds lift,''
  arXiv:hep-th/0603066.

\bibitem{0605210}
  J.~R.~David and A.~Sen,
  ``CHL dyons and statistical entropy function from D1-D5 system,''
  JHEP {\bf 0611}, 072 (2006)
  [arXiv:hep-th/0605210].

\bibitem{0607155}
  J.~R.~David, D.~P.~Jatkar and A.~Sen,
  ``Dyon spectrum in N = 4 supersymmetric type II string theories,''
  arXiv:hep-th/0607155.


\bibitem{0609109}
  J.~R.~David, D.~P.~Jatkar and A.~Sen,
  ``Dyon spectrum in generic N = 4 supersymmetric Z(N) orbifolds,''
  arXiv:hep-th/0609109.

\bibitem{0612011}
  A.~Dabholkar and D.~Gaiotto,
  ``Spectrum of CHL dyons from genus-two partition function,''
  arXiv:hep-th/0612011.

\bibitem{0702141}
  A.~Sen,
  ``Walls of marginal stability and dyon spectrum in N = 4 supersymmetric
  string theories,''
  arXiv:hep-th/0702141.
  
\bibitem{0702150}
  A.~Dabholkar, D.~Gaiotto and S.~Nampuri,
  ``Comments on the spectrum of CHL dyons,''
  arXiv:hep-th/0702150.
  
\bibitem{0705.1433}
  N.~Banerjee, D.~P.~Jatkar and A.~Sen,
  ``Adding charges to N = 4 dyons,''
  arXiv:0705.1433 [hep-th].

\bibitem{0705.3874}
  A.~Sen,
  ``Two Centered Black Holes and N=4 Dyon Spectrum,''
  arXiv:0705.3874 [hep-th].

\bibitem{0706.2363}
  M.~C.~N.~Cheng and E.~Verlinde,
  ``Dying Dyons Don't Count,''
  arXiv:0706.2363 [hep-th].

\bibitem{0005049}
  F.~Denef,
  ``Supergravity flows and D-brane stability,''
  JHEP {\bf 0008}, 050 (2000)
  [arXiv:hep-th/0005049].

\bibitem{0010222}
F.~Denef,
``On the correspondence between D-branes 
and stationary supergravity solutions of type
II Calabi-Yau compactifications'', 
arXiv:hep-th/0010222.

\bibitem{0101135}
  F.~Denef, B.~R.~Greene and M.~Raugas,
  ``Split attractor flows and the spectrum 
of BPS D-branes on the 
quintic,''
  JHEP {\bf 0105}, 012 (2001)
  [arXiv:hep-th/0101135].

\bibitem{0304094}
  B.~Bates and F.~Denef,
  ``Exact solutions for supersymmetric stationary black hole 
composites,''
  arXiv:hep-th/0304094.

\bibitem{0702146}
  F.~Denef and G.~W.~Moore,
  ``Split states, entropy enigmas, holes and halos,''
  arXiv:hep-th/0702146.

\bibitem{0705.2564}
  F.~Denef and G.~W.~Moore,
  ``How many black holes fit on the head of a pin?,''
  arXiv:0705.2564 [hep-th].
  
  \bibitem{denef}
  F.~Denef, private communications.
  
  \bibitem{cheng}
  M.~Cheng, private communications.

\bibitem{9507090}
  M.~Cvetic and D.~Youm,
  ``Dyonic BPS saturated black holes of heterotic string on a six torus,''
  Phys.\ Rev.\ D {\bf 53}, 584 (1996)
  [arXiv:hep-th/9507090].
 
 \bibitem{9508094}
  M.~J.~Duff, J.~T.~Liu and J.~Rahmfeld,
  ``Four-Dimensional String-String-String Triality,''
  Nucl.\ Phys.\ B {\bf 459}, 125 (1996)
  [arXiv:hep-th/9508094].
  

\end{thebibliography}
\end{document}